\documentclass[10pt]{article}

\usepackage{amsmath}
\usepackage{amssymb}

\usepackage{graphicx}

\usepackage{cite}

\usepackage{color} 

\usepackage[labelfont=bf,labelsep=period,justification=raggedright]{caption}

\makeatletter
\renewcommand{\@biblabel}[1]{\quad#1.}
\makeatother

\date{}

\newcommand{\red}{\textcolor{black}}
\newcommand{\blue}{\textcolor{black}}

\begin{document}

\begin{flushleft}
{\Large
\textbf{Sequence alignment, mutual information, and dissimilarity measures for constructing phylogenies}
}
\\
\vglue .8cm
Orion Penner$^{1,\ast}$, 
Peter Grassberger$^{1,2,\ast}$, 
Maya Paczuski$^{1,\ast}$
\\
\vglue .3cm
\bf{1} Complexity Science Group, Department of Physics and Astronomy, University of Calgary, Calgary, Alberta, Canada
\\
\bf{2} Institute for Biocomplexity and Informatics, Department of Biological Sciences, University of Calgary, Calgary, Alberta, Canada
\\
$\ast$ E-mail: openner@phas.ucalgary.ca, pgrassbe@ucalgary.ca, maya.paczuski@ucalgary.ca
\end{flushleft}

\section*{Abstract}
\paragraph*{Background:} Existing sequence alignment algorithms use heuristic scoring schemes based on biological expertise, which cannot be used as objective distance metrics. As a result one relies on crude measures, like the p- or log-det distances, or makes explicit, and often too simplistic, {\it a priori} assumptions about sequence evolution. Information theory provides an alternative, in the form of mutual information (MI). MI is, in principle, an objective and model independent similarity measure, but it is not widely used in this context and no algorithm for extracting MI from a given alignment (without assuming an evolutionary model) is known. MI can be estimated without alignments, by concatenating and zipping sequences, but so far this has only produced estimates with uncontrolled errors, despite the fact that the {\it \red{normalized} compression distance} based on it has shown promising results.

\paragraph*{Results:} We describe a simple approach to get robust estimates of MI from {\it global pairwise} alignments. Our main result uses algorithmic (Kolmogorov) information theory, but we show that similar results can also be obtained from Shannon theory. For animal mitochondrial DNA our approach uses the alignments made by popular global alignment algorithms to produce MI estimates that are strikingly close to estimates obtained from the alignment free methods mentioned above. We point out that, due to the fact that it is not additive, \red{normalized} compression distance is not an optimal metric for phylogenetics but we propose a simple modification that overcomes the issue of additivity. We test \blue{several versions of our MI based distance measures on a large number of randomly chosen quartets and demonstrate that \blue{they all perform better than traditional measures like the Kimura or } log-det (resp. paralinear) distances.}

\paragraph*{Conclusions:} \blue{Several versions of MI based distances outperform conventional distances in distance-based phylogeny. Even a simplified version based on single letter Shannon entropies, which can be easily incorporated in existing software packages, gave superior results throughout the entire animal kingdom.} But we see the main virtue of our approach in a more general way. For example, it can also help to judge the relative merits of different alignment algorithms, \blue{by} estimating the significance of specific alignments. It strongly suggests that information theory concepts can be exploited further in sequence analysis.

\section*{Background}

Sequence alignment achieves many purposes and comes in several different varieties \cite{aluru2006}: Local versus global (and even ``glocal'':~\cite{brudno2003gaf}), pairwise versus multiple, and DNA/RNA versus proteins. Rather than listing all applications, we cite just two numbers: \blue{According to Google Scholar the two original papers on the BLAST algorithm for local alignment by \cite{altschul1990} and on one of its improvements \cite{altschul1997} have been cited more than 30,000 times each}, and the number of daily file uploads to the NCBI server providing BLAST is $\approx 140,000$ \cite{mcginnis2004bcp}. A partial list of alignment tools in the public domain can be found in {\sf http://pbil.univ-lyon1.fr/alignment.html}.

In {\it global} alignment, which we focus on here, two sequences of comparable length are placed one below the other.  The algorithm inserts blanks in each of the sequences such that the number of positions at which the two sequences agree is maximized.  More precisely, a {\it scoring scheme} is used.  Each position at which the two sequences agree is rewarded by a positive score, while each disagreement (``mutation'') and each insertion of a blank (``gap'') is punished by a negative one.  The best alignment is that with the highest total score.  In {\it local} alignment, one aligns only subsequences against each other and looks for the highest scores between any pairs of subsequences. Regions that cannot be well-aligned are simply ignored.  Existing algorithms use either heuristic scoring schemes or scores derived from explicit probabilistic models \cite{durbin1998bsa}.

Similarities between DNA sequences, e.g. for distance-based phylogenetic tree construction, are typically not based on alignment scores.  Instead they use explicit evolutionary assumptions (e.g. the Kimura two-parameter model \cite{Nei-Kumar2000}) or are simply obtained by counting the number of nucleotide substitutions (like the p-distance or the Poisson corrected p-distance \cite{Nei-Kumar2000}). An important property of a similarity measure, from the point of view of phylogeny, is that distances should grow linearly with evolution time.  This results in a measure satisfying the so-called {\it four point condition} \cite{Buneman1974}, which in turn makes the measure useful for {\it neighbor joining}, the most popular distance based algorithm for inferring phylogenetic trees \cite{Saitou-Nei}. The most important metrics developed from this view point are the closely related {\it paralinear} \cite{Lake1994} and {\it log-det} \cite{Lockhart1994} distances.  In this paper we refer to both as ``log-det", for simplicity's sake.

In the above mentioned distances, distinct rates of different substitution types are either taken into account using a model, or are not taken into account at all \footnote{The case of the log-det distance is somewhat subtle. There, substitution types are taken into account in the definition, but the definition is made such that the distance depends only on the expected total number of substitutions in a Markov model, provided the substitution matrix is reversible.  \blue{Similarly, in models like the 2-parameter Kimura model substitution types are taken into account when esimating parameters, but the actual distance is an estimated unweighted sum over all substitutions.} }. This fact stands in stark contrast with {\it mutual information} (MI), \blue{which} takes the amount of information shared between two objects as a measure of their similarity \cite{cover}. \blue{For instance,} more frequent substitutions can be encoded more efficiently, and should thus be a weaker indicator for dissimilarity than rare, and thus ``surprising", substitutions. The crucial point to note is that the frequency of substitutions \blue{and indels and their correlations} can be counted directly from the alignment, and no model is required. As a consequence, MI is\red{, in principle,} a model-free, universal, and objective similarity measure, in stark contrast to all metrics discussed above.

Indeed, there are two variants of information theory: The more traditional Shannon theory, based on a probabilistic interpretation of the sequences, and the less well known Kolmogorov (or {\it algorithmic}) ``complexity" theory \cite{cover}. In this paper we use Kolmogorov information as our main vehicle, \blue{but we also show that} Shannon theory \blue{gives comparable results}. 

Roughly, in algorithmic information theory the {\it complexity} $K(A)$ of a sequence $A$ is the minimal amount of information (measured in bits) needed to specify $A$ uniquely, on a given computer, with a given operating system. Numerical results depend on the latter, but this dependence will, in general, be weak and \blue{is} ignored in this paper. For two sequences $A$ and $B$, the conditional complexity (or conditional information) $K(A|B)$ is the information needed to specify $A$, if $B$ is already known (i.e., either it or its specification was already input before). If $A$ and $B$ are similar, this information might consist of a short list of changes needed to go from $B$ to $A$, and $K(A|B)$ is small. If, on the other hand, $A$ and $B$ have nothing in common, then knowing $B$ is useless and $K(A|B)=K(A)$. Finally, the {\it mutual information} (MI) is defined as the difference $I(A;B) = K(A)-K(A|B).$  It is the amount of information which is common to $A$ and $B$, and is also equal to the amount of information in $B$ which is useful for describing $A$, and {\it vice versa}. Indeed, it can be shown that, up to correction terms that become negligible for long sequences (see~\cite{cover}): (a) $I(A;B) \geq 0$; (b) $I(A;B) = 0$ if and only if $A$ and $B$ are completely independent; (c)  $I(A;A) = K(A)$; and (d) $I(A;B) = I(B;A)$. Moreover, the likelihood that  $A$ and $B$ \blue{were generated} independently is  $p = 2^{-I(A;B)}$ (see~\cite{milos95}).  Hence, the similarity is significant and not by chance when $I(A;B)$ is large.

The fundamental difference between Shannon theory and Algorithmic Information theory is that Shannon theory makes no attempt to quantify the minimum information required to specify a any particular sequence.  Instead Shannon theory assumes that a sequence can be treated as though it was generated as a "typical" case of a probabilistic process.  As a result of this assumption, Shannon information has no dependence on hardware. However, the main drawback is that it cannot, strictly speaking, deal with individual sequences \blue{and it needs an assumption on the probability distribution}. Any numerical result obtained from individual sequences implies the assumption that the specific sequences are `typical' of the underlying probabilistic process. As a result it involves statistical inference, even if the result does not strongly depend on this inference. \blue{ In particular, the assumption of independence of letters in a sequence (used also below) will lead to over-estimation of Shannon entropy, and thus implies no risk of overfitting.}  

The fact that alignment and information theory are closely related has been realized repeatedly.  However, most work in this direction has focused on aligning images rather than sequences \cite{viola1997}. Conceptually, these two problems are closely related, but technically, they are not. The effects of sequence randomness on the significance of alignments has also been studied in \cite{Allison99}.  Finally, attempts to extend the notion of {\it edit distance} \cite{aluru2006} to more general editing operations have been made. In this case the similarity of two sequences is quantified by the complexity of the edit string, see \cite{varre99}.  Indeed, the aims of \cite{varre99} are similar to ours, but their approach differs in several key respects and leads to markedly different results.

\section*{Methods}
\subsection*{Translation Strings}

\begin{figure}
\centering
\includegraphics[width=0.75\textwidth]{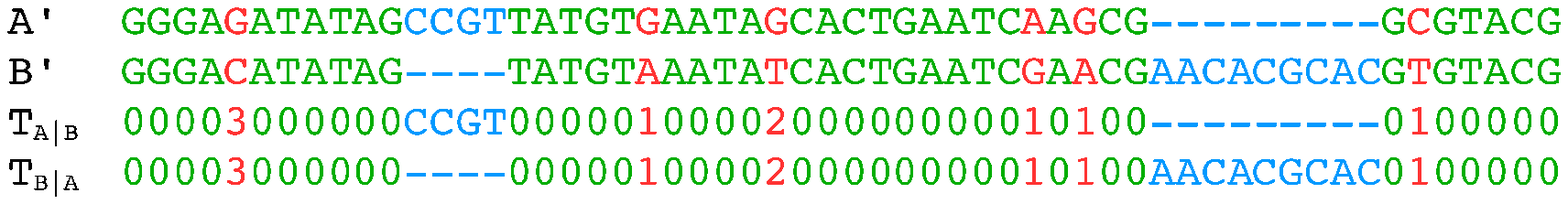}
\caption{Example of an alignment and of the two translation strings $T_{A|B}$
  and $T_{B|A}$. Colors indicate sites with mutations (red), gaps (blue), and 
  conservation (green).
  \label{fig:example_trans}}
\end{figure}

At the heart of our approach is the concept of a {\it translation string}.  The 
translation string $T_{B|A}$ contains the minimal information necessary to recover 
the sequence $B$ from another sequence $A$. Similarly, $T_{A|B}$ contains the 
information needed to obtain $A$ from $B$.  Here we focus on  DNA sequences, 
consisting of the letters A,C,G and T, and  corresponding to complete mitochondrial 
genomes.  But the approach is more general and can be applied to protein sequences 
without further effort.  We refer to the $i^{th}$ element of sequence $X$ as $X_i$, 
and denote the length of $X$ as $n_X$.  Any global alignment algorithm, when applied 
to $A$ and $B$, outputs a pair of sequences $(A',B')$ of equal length $n\geq \max\{n_A,n_B\}$. 
The sequences $A'$ and $B'$ are obtained from $A$ and $B$ by inserting hyphens 
("gaps") such that the total score is maximized. The strings $T_{B|A}$ and $T_{A|B}$ 
also have length $n$, and are composed from an alphabet of nine characters. For each 
$i$, the letter $T_{B|A,i}$ is a function of $A'_i$ and $B'_i$ only. An example of 
this process is found in Figure~1; the rules to create $T_{B|A}$ are as follows:
\begin{itemize}
\item if $A'_i = B'_i$, then $T_{B|A,i}=0$;
\item if $A'_i$ is a hyphen (gap), then $T_{B|A,i}$ has to specify explicitly what 
is in $B$; hence $T_{B|A,i} = B'_i \in\{$A,C,G,T$\}$;
\item if $B'_i$ is a hyphen (gap), then $T_{B|A,i}$ has to indicate that {\it something}
is deleted from $A'$, but there is no need to specify what. Hence $T_{B|A,i} = B'_i = -$;
\item if $A'_i \to B'_i$ is a {\it transition}, i.e. a substitution A$\leftrightarrow$G 
        or C$\leftrightarrow$T, then $T_{B|A,i}=1$;
\item if $A'_i \to B'_i$ is a {\it transversion} A$\leftrightarrow$C or T$\leftrightarrow$G,
        then $T_{B|A,i}=2$;
\item if $A'_i \to B'_i$ is a transversion A$\leftrightarrow$T or G$\leftrightarrow$C,
        then $T_{B|A,i}=3$.
\end{itemize}
$T_{B|A}$ is defined such that $B'$ (and thus also $B$) is obtained uniquely from $A'$.  But $A'$ can be obtained from $A$ using $T_{B|A}$. Thus $T_{B|A}$ does exactly what it is intended to do: it allows one to recover $B$ from $A$. It does not, however, allow one to recover $A$ from $B$. Due to the second and third bullet points above, $T_{B|A}$ is not the same as $T_{A|B}$.  This distinguishes our approach from typical edit string methods.

\subsection*{Algorithmic Information Theory: Mutual Information}

An estimate of the conditional complexity $K(B|A)$ is obtained by compressing $T_{B|A}$ using any general purpose compression algorithm 
such as zip, gzip, bzip2, etc. In the results shown here we use lpaq1 \cite{mahoney} (see also this reference for a survey of public 
domain lossless compression algorithms). Denoting by ${\rm comp}(A)$ the compressed version of $A$ and by ${\rm len}[A]$ the length 
of $A$ in bits, gives an exact upper bound
\begin{equation}
\blue{
   K(B|A) \leq K_0(B|A) \equiv {\rm len}[{\rm comp}(T_{B|A})].                            \label{Kest0}
}
\end{equation}
If there were no correlations between sequence $A$ and the translation string, $T_{B|A}$, 
this would also be the best possible upper bound. However, in general we must expect that 
such correlations exist, although we find them to be weak (see Figure S2 in the supplementary 
material). Thus Eq.~(\ref{Kest0}) is still a good estimate, but the best one is obtained by 
compressing conditionally on $A$, 
\begin{equation}
   K(B|A) \leq {\rm len}[{\rm comp}(T_{B|A}|A)]   
        \equiv  {\rm len}[{\rm comp}(A\;T_{B|A})] -  {\rm len}[{\rm comp}(A)] \leq  K_0(B|A) .                                             \label{Kest}
\end{equation}
More precisely, one can show that $K_0(B|A)-K(B|A) \approx I(A;T_{B|A})$.

In order to obtain an estimate of  MI, we have to subtract $K(B|A)$ from $K(B)$, which is also 
estimated via compression. Unlike $T_{B|A}$,  $B$ is a DNA string. Since general purpose 
compression algorithms are known to be inferior for DNA \cite{gencomp,cao2007ssa} we could 
use an efficient DNA compressor like, e.g., `GeneCompress" or ``XM" \cite{gencomp,cao2007ssa} 
\blue{(as we shall do in Eq.~(\ref{I_compr}) below)}. To avoid any question of consistency, we shall 
not do this. \blue{Instead,} the compression is carried out using a general purpose compression 
algorithm, \blue{to get}
\begin{eqnarray}
   I(A;B) \approx I(A;B)_{\rm align} &=& {\rm len}[{\rm comp}(B)] - {\rm len}[{\rm comp}(T_{B|A}|A)] \nonumber \\
          &=& {\rm len}[{\rm comp}(A)] +  {\rm len}[{\rm comp}(B)] - {\rm len}[{\rm comp}(A\;T_{B|A})].
                           \label{I_align}
\end{eqnarray}

This is to be compared to the general definition of algorithmic MI, based entirely
on concatenation and compression \cite{cover,cili2005} \blue{without using any alignment. 
This estimate} is obtained by comparing the size of the compressed concatenation $AB$ to 
the sum of the sizes of the compressed individual files,
\begin{equation}
   I(A;B)_{\rm compr} = {\rm len}[{\rm comp}(A)] + {\rm len}[{\rm comp}(B)]
                      - {\rm len}[{\rm comp}(AB)].                 \label{I_compr}
\end{equation}

At first sight it might seem paradoxical that $I(A;B)_{\rm align}$ can even be
positive.  Not only does $T_{B|A}$ involve a larger alphabet than $B$, but, in
general, it is also a longer string. Thus one could expect that $T_{B|A}$ would
not typically compress to a shorter size than $B$. The reason why this first
impression is wrong is clear from Figure~1: If $A$ and $B$
are similar, then $T_{B|A}$ consists mostly of zeroes and compresses readily. In 
practical alignment schemes, the scores for mismatches are carefully chosen such 
that more frequent substitutions are punished less than unlikely substitutions. In
contrast, coding each mismatch simply by a letter in $T_{B|A}$ seems to ignore
this issue.  However, more frequent mismatches will give letters occurring with
higher frequency, and general purpose compression algorithms utilize frequency
differences to achieve higher compression.

Conceptually our approach is similar to encoding generalized edit strings
in \cite{varre99}. However, there are several pivotal differences between that
work and ours.  First, the authors in \cite{varre99} did not compress their
edit strings and as a result the conclusions they were able to draw from
a quantitative analysis were much weaker than ours.  Second, our approach
utilizes an alignment algorithm to achieve an efficient encoding of $T_{B|A}$.
In addition to producing a better estimate of $K(B|A)$, this allows us to make
quantitative evaluations of the alignment algorithm itself.  An additional difference
between our approach and the traditional edit methods used in approximate
string matching \cite{navarro2001}  is that our {\it translation strings} do not
give both translations $A\to B$ and $B\to A$ from the same string.  This
asymmetry is crucial to establish the relations to conditional and mutual information.

For long strings, $I(A;B)$ should be symmetric in its arguments. 
In general, the estimates satisfy $I(A;B)_{\rm align}\approx I(B;A)_{\rm align}$
(see Figure~S3 in the supplementary material). Indeed, the translation 
strings $T_{B|A}$ and $T_{A|B}$ can differ substantially, resulting in 
different estimates for $K(B|A)$ and $K(A|B)$ via Eq.~(\ref{Kest}). This
difference is mostly canceled by differences between ${\rm len}[{\rm comp}(B)]$ 
and ${\rm len}[{\rm comp}(A)]$. Take, for instance, the case
where $B$ is much shorter than $A$. Then $T_{B|A}$ consists 
mostly of hyphens and is highly compressible. On the other hand, $T_{A|B}$
is similar to $A$, since most letters have to be inserted when translating 
$B$ to $A$. Thus both $I(A;B)_{\rm align}$ and $I(B;A)_{\rm align}$ are 
small compared to $K(A)$, but for different reasons. Further details are 
given in the supplementary material.

\subsection*{Shannon Theory}

Compared to algorithmic information, Shannon theory is the more widely known version 
of information theory \cite{cover}. The basic concept of Shannon theory is that of a 
{\it block} or {\it word} probability $p_n(s_1\ldots s_n|A)$. It gives the probability 
that the `word' $s_1\ldots s_n$ of $n$ consecutive letters (such as A,C,G or T for DNA) 
appears at any random position in the string $A$. Here we assume stationarity, but we 
do not assume absence of correlations. The entropy (analogous to the complexity in 
algorithmic information theory) of a string comprised of letters from an alphabet 
${\cal A}$ is defined as $h(A) = \lim_{n\to \infty} h_n(A)$ with
\begin{equation}
   h_n(A) = - \sum_{\{s_1\ldots s_n\} \in {\cal A}^n}
               p_n(s_1\ldots s_n|A) \log p_n(s_1\ldots s_n|A). 
\end{equation}
From this, MI is defined as in algorithmic theory: $I(A;B) = h(A)+h(B)-h(A,B)$ \cite{cover}. 
If entropy is measured in bits, then the logarithm is to base 2. In practice, the limit 
$n\to\infty$ is rarely feasible, and one usually approximates $h(A)$ by the single-letter entropy
\begin{equation}
   h(A) \approx h_1(A) \equiv - \sum_{s\in \cal A} p(s) \log p(s)      \label{h1}
\end{equation}
or, at most, by the pair approximation based on the probabilities for words of length two. 

Eq. (\ref{h1}) is valid under the assumption that correlations between consecutive letters in 
the string can be neglected. \blue{Similarly, $h(A,B)$ for two sequences of equal length is estimated 
by assuming that consecutive letter pairs $(s_i,t_i)$ with $s_i\in A$ and $t_i\in B$ are independent}. 
If we make this assumption, there are still two ways to estimate the MI of two strings. In the first we 
use \blue{the fact that $(A',B')$ carries the same information as $(A,B)$ to employ a five-letter 
alphabet ${\cal A} = \{A,C,G,T,-\}$. This has the drawback that indels are usually correlated.} In the 
second we thus neglect all indels and reduce the alphabet to ${\cal A} = \{0,1,2,3\}$. In the following 
we shall mostly use the latter to compare with other pairwise distance metrics, but we stress that 
we do this only for simplicity and convenience (and since it \blue{is} sufficient to make our point). 
However, the more interesting MI estimate remains the one obtained from algorithmic theory, due to the 
fact it takes into account both indels and all possible correlations within each string and between 
\blue{them}.

\subsection*{Distances, Trees and Quartets}

The value of the MI itself is useful for many purposes\blue{: E}stimating similarities between different 
pairs (and thus of finding closest neighbors of a given sequence in a large data set); comparing the qualities 
of alignments obtained by different algorithms; or assessing the significance of an alignment (i.e., verifying 
that it is better than an alignment between two unrelated sequences). But in the case of phylogeny, one wants 
more. Ideally, one wants an {\it additive metric distance}, i.e. a non-negative symmetric pairwise function 
$d_{AB}$ for which $d_{AA}=0$ and which satisfies both the triangle inequality 
\begin{equation}
   d_{AC}\leq d_{AB} + d_{BC} \label{additive}
\end{equation}
for any triple, and the {\it four-point condition} \cite{Press2007} 
\begin{equation}
   d_{AB} + d_{CD} \leq \max \{d_{AC} + d_{BD},d_{AD}+d_{BC}\}.     \label{4-pt}
\end{equation}
for any quartet.
The latter is a necessary and sufficient condition for all pairwise distances between $N$ sequences to be 
representable as distance sums over links in a tree \cite{Buneman1974} with the $N$ sequences represented 
by the leaves. \blue{Thus} distances satisfying Eq.~(\ref{4-pt}) are also called `tree metrics'. 
 
Several potential metrics can be derived from MI \cite{li2001ibs,li2004,cili2005}. According 
to \cite{li2004,cili2005}, the preferred one is the {\it \red{normalized} compression distance}
\begin{equation}
    d^{\rm(\red{NCD})}_{AB} = 1 - {I(A;B)\over \max\{K(A),K(B)\}}.                    \label{NCD}
\end{equation}
For Shannon theory we can use the same construct with $K(A)$ replaced by $h(A)$ \cite{Kraskov2003}. Since 
it would be confusing to use the word ``compression" \blue{for this metric}, we have to use another name. We call 
it the {\it \red{normalized} Shannon distance}
\begin{equation}
    d^{\rm(\red{NSD})}_{AB} = 1 - {I(A;B)\over \max\{h(A),h(B)\}}.                    \label{NSD}
\end{equation}

Although $d^{\rm(\red{NCD})}$ has been used to produce meaningful phylogenetic trees 
\cite{li2001ibs,li2004,cili2005,Kraskov2003}, it has one important drawback for phylogenetic 
applications: It is not additive. Indeed, for two completely unrelated sequences 
(corresponding to infinite evolutionary distance), both $d^{\rm (\red{NCD})}_{AB}$ 
and $d^{\rm (\red{NSD})}_{AB}$ do not go to infinity, but rather to 1.  They are not 
linear but convex functions of evolutionary distance. Such metrics are well known to lead 
to {\it long branch attraction} (or \blue{the} `Felsenstein phenomenon' \cite{Felsenstein1981}).

If evolution is assumed to be a Markov process, then the data processing inequality \cite{cover}
guarantees that MI decreases with evolutionary distance. A natural assumption \blue{-- following from 
the dominance of a single maximal eigenvalue of the Markov matrix --} is 
that it decreases exponentially to zero. In \blue{this} case the {\it log-MI} ``distance"
\begin{equation}
      d^{\rm(log-MI)}_{AB} = - \log \left[{I(A;B)\over \max\{K(A),K(B)\}}\right]
      \quad {\rm resp.}\quad - \log \left[{I(A;B)\over \max\{h(A),h(B)\}}\right]       \label{logdist}
\end{equation}
would increase linearly with evolution and would be thus additive. Unfortunately, 
$d^{\rm(log-MI)}_{AB}$ is not a proper metric, as it does not even satisfy the 
triangle inequality.  This can be seen from the following example: Take three sequences 
over an alphabet of four letters (like DNA) where each letter is represented by two 
bits (purine/pyrimidine, double/triple hydrogen bonds). Sequence $B$ is random, sequence 
$A$ is obtained from $B$ by replacing randomly the first bit but conserving the second, 
and $C$ is obtained by replacing the second but conserving the first. Then $I(A;B)$ and
$I(B;C)$ are non-zero, while $I(A;C)=0$. At the same time, all single sequence 
complexities (and entropies) are the same, thus 
$d^{\rm(log-MI)}_{AC} = \infty$ while $d^{\rm(log-MI)}_{AB}$ and $d^{\rm(log-MI)}_{BC}$
are finite, clearly violating Eq. (\ref{additive}).

Fortunately, real evolution is most likely not as extreme as this counter example, 
and the triangle inequality is not really required for distance based phylogeny. In 
particular, the relationship between trees and metric additivity is not restricted to 
metrics satisfying the triangle inequality, as seen from the proof in \cite{Buneman1974}.
Also, the neighbor joining algorithm \cite{Saitou-Nei} does not require the triangle 
inequality. Thus we claim that $d^{\rm(log-MI)}$ is an {\it a priori} better distance 
measure for phylogeny than $d^{\rm(\red{NCD})}$ or $d^{\rm(\red{NSD})}$, although a
final evaluation can only be made through detailed tests on real biological sequences.

Such tests are presented in the results section, with the log-det (or, more precisely, 
the paralinear) distance \cite{Lake1994,Lockhart1994} \blue{and two distances based on 
Kimura's model \cite{Nei-Kumar2000} (see the supplementary information) as other 
competitors}. In the latter, one assumes different rates $\alpha$ for transitions 
(A$\leftrightarrow$ G, C$\leftrightarrow$T) and $\beta$ for transversions (all others).

Assume that for two aligned sequences, $A$ and $B$, one first eliminates all positions 
with indels.  Thus, at each site one sees one of the 16 possibilities $(i,k)$ with 
$i,k \in \{A,C,G,T\}$.  Denote the measured frequencies for these possibilities $f_{AB}(i,k)$. 
The single-sequence (`marginal') frequencies are $g_A(i) = \sum_k f_{AB}(i,k)$ and 
$g_B(k) = \sum_i f_{AB}(i,k)$. We introduce matrices $F_{AB}$ with matrix elements
$(F_{AB})_{ik}= f_{AB}(i,k)$, $G_A$ with $(G_A)_{ik} = g_A(i)\delta_{ik}$, and 
$(G_B)_{ik} = g_B(i)\delta_{ik}$ (here, $\delta_{ik}$ is the Kronecker delta, i.e. $G_A$
and $G_B$ are diagonal matrices). The log-det distance is then defined as
\begin{equation}
      d^{\rm(log-det)}_{AB} = - \log\; \det [G_A^{-1/2} F_{AB} G_B^{-1/2}]. \label{logdet}
\end{equation}
In \cite{Lake1994}, this is called paralinear distance; in \cite{Lockhart1994} the name 
log-det is used either for this or for simplified versions where the matrices $G_A$ 
and $G_B$ are omitted. This difference is irrelevant for additivity and \blue{for use} in the 
neighbor joining algorithm. It can be shown that $d^{\rm(log-det)}_{AB}$ is additive 
under rather general evolutionary models, although not when evolutionary speed is 
site dependent. 

Before moving on, we should point out that the data required to compute
the log-det distance are {\it precisely} the same as those 
needed to compute the two MI-based distances $d^{\rm(\red{NSD})}_{AB}$ and 
$d^{\rm(log-MI)}_{AB}$, provided one uses for the latter the single-letter Shannon
formulas with indels deleted. In that case,
\begin{equation}
     I(A;B) = \sum_{i,k} f_{AB}(i,k) \log{f_{AB}(i,k)\over g_A(i) g_B(k)} \label{I-shannon}
\end{equation}
and 
\begin{equation}
    h(A) = -\sum_i {g_A(i)\over M} \log {g_A(i)\over M}                \label{h-shannon}
\end{equation}
where $M = \sum_i g_A(i) = \sum_k g_B(k)$, and is the number of sites in the alignment.
This is the main reason we will later compare these three distances in detail.

\subsection*{Tools}
We utilized the MAVID \cite{bray2003mma} and Kalign \cite{lassmann2005kaa} global 
sequence alignment programs available for download at \cite{mavid} and \cite{kalign}.
We also experimented with STRETCHER \cite{stretcher},
lagan \cite{brudno2003lam} and CLUSTALW 2 \cite{thompson22gci}, and observed 
similar results. We made no effort to optimize the scoring parameters of the 
algorithms and only used the default \blue{values}.

\blue{To evaluate $I(A;B)_{\rm compr}$} we utilized the {\it expert model} (XM) DNA compression 
algorithm \cite{cao2007ssa}. \blue{To evaluate $I(A;B)_{\rm align}$} we used 
lpaq1 \cite{mahoney}. Using lpaq1 was not crucial, with the standard LINUX 
tools gzip and bzip2 producing similar results. For DNA we also explored
GenCompress \cite{li2001ibs} and bzip2. Both showed markedly inferior results to XM 
(see supplementary information), \blue{although their ability to compress single sequences 
is not so much inferior to XM \cite{cao2007ssa}. Presumably this is due to the fact that 
XM is more efficient in finding and exploiting approximate repeats, which is crucial 
in compressing concatenated strings}.

The complete mtDNA sequences used in our analysis were downloaded from \cite{ncbi}. We paid
special attention to eliminate incomplete sequences and sequences with too many wild cards. 
We also took care to circularly shift the sequences (mtDNA forms in most cases a 
closed ring) in order to improve the alignments. We used different subsets of 
sequences for different plots. In a few cases we also flipped the strands, if this 
led to much better alignments. Overall, we used nearly 1800 sequences.

\section*{Results}
\subsection*{Alignment based mutual informations versus compression based mutual informations}

\begin{figure}[t]
\centering
\includegraphics[width=0.45\textwidth,angle=270]{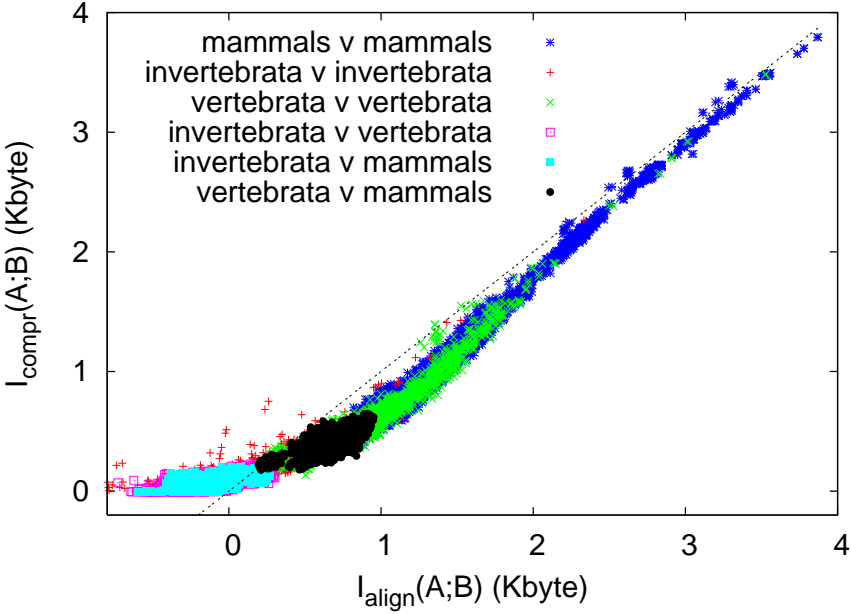}
\vglue 8.mm
\caption{Scatter plot of MI estimates for complete mitochondrial DNA between pairs of
  species: $I_{\rm compr}$ using XM \cite{cao2007ssa} vs. $I_{\rm align}$ using 
  MAVID \cite{bray2003mma} followed by compression with lpaq1. Note that the two estimates
  generally agree and fall on the diagonal, while in some cases one method does better 
  than the other as explained in the text. Here and in Figure~3 
  ``vertebrata'' means non-mammalian vertebrata. This plot contains roughly 36,000
  pairs, about 16,000 of which contain two mammals, the other 20,000 covering 
  equally the other combinations. $I_{\rm align}$ is the 
  average between the values obtained from $T_{B|A}$ and $T_{A|B}$.}
  \label{fig:mavid_PG_vs_XM}
\end{figure}

Our first results concern the agreement between the two estimates $I_{\rm align}$ and 
$I_{\rm compr}$. \blue{In Figure~\ref{fig:mavid_PG_vs_XM} we compare estimates 
$I_{\rm compr}$ obtained with XM to estimates $I_{\rm align}$ obtained
with the MAVID alignment tool \cite{mavid} and with subsequent compression using
lpaq1.} It is well known that DNA and amino acid 
sequences are hard to compress \cite{gencomp,cao2007ssa}, thus one might expect 
that $I_{\rm compr}$ depends strongly on the compression algorithm used. This is indeed 
the case, as seen from Figure S1 in the supplementary material, where we compare values 
of $I_{\rm compr}$ obtained with three different compression algorithms: The general 
purpose compressor lpaq1 \cite{mahoney} and the two special DNA compressors GeneCompress
\cite{gencomp} and XM \cite{cao2007ssa}. From this figure it is clear that XM is far 
better the other two. \blue{Note that} it is very likely that an imperfect compression 
algorithm underestimates \blue{rather than overestimates MI} -- although we do not know 
a rigorous theorem to this effect. 

In view of this, it is not obvious that the estimates produced by 
XM are realistic either. \blue{It is thus highly significant that the two estimates 
shown in Figure~\ref{fig:mavid_PG_vs_XM}} are
approximately equal, despite the fact that alignment algorithms and compression
algorithms follow drastically different routes. The slight downward shift from the 
diagonal, particularly visible for large MI pairs, is due to an off-set of 
$\approx 50$ bytes in the XM algorithm. Points above the diagonal
indicate that concatenation and compression -- using the XM algorithm --
produce a better estimate of MI, while points below indicate that MAVID
alignment followed by compression of its translation string produced a better
estimate. The invertebrate-invertebrate pairs far above the diagonal in
Figure~2 correspond  to pairs of species where the individual 
genes are  similar, but their ordering is changed (this refers in particular to all 
pairs with $I_{\rm align}\approx 0$ and $I_{\rm compr} > 0.3$ Kbyte). 
In that case a compression algorithm is superior to a global alignment algorithm, 
since it is not affected by shuffling open reading frames (ORFs). Most negative
estimates for  MI seen in Figure~2 represent cases where
shuffling the ORFs prevented reasonable global alignments. Particularly interesting 
are pairs of mammals with $I_{\rm compr} > I_{\rm align}$. We checked that {\it all} of 
them involve a subspecies of sikka deer (Cervus nippon taiouanus, GenBank accession 
number DQ985076), in which a single gene (NADH6) is supposedly on the opposite strand 
compared to all other mammals. 

\begin{figure}
\centering
\includegraphics[width=0.68\textwidth]{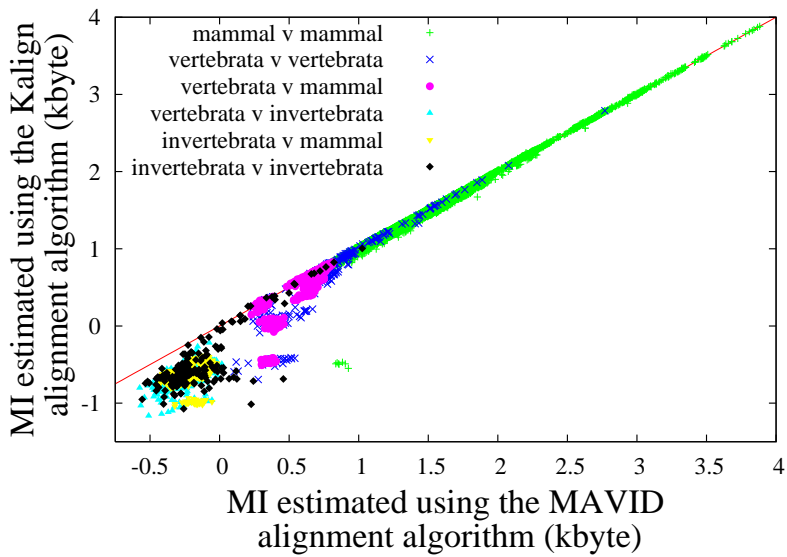}
\caption{Scatter plot comparing alignment based MI estimates:
  Kalign \cite{lassmann2005kaa} vs. MAVID \cite{bray2003mma}. The number of pairs here 
  is about half of that shown in Figure~2. Points on the diagonal indicate agreement between 
  the two estimates. These data were generated using the default scoring parameters. 
  Therefore, the plot represents a proof of principle \blue{for using MI to evaluate alignments}
  rather than a definitive statement about the quality of the two alignment algorithms shown.
\label{fig:kalign_PG_vs_mavid_PG}
}
\end{figure}

Agreement between $I_{\rm compr}$ and $I_{\rm align}$ could have been improved \blue{presumably}
in many cases by masking part of the genome, but we have not tried this. In any 
case, the occasional disagreements are of particular interest, since 
they indicate where one of the two approaches encountered particular difficulty.
Generally speaking Figure~2 suggests that DNA compression can still be improved 
slightly, as seen from pairs with $I$ between 1 and 2 Kbyte (corresponding roughly 
to species in different families but the same orders). On the other hand, purely 
compression based MI estimates give non-trivial \blue{(at least positive)} results even 
across different classes. 

\subsection*{Comparison between different alignment algorithms}

MI estimates obtained using other global  alignment algorithms are  similar to those 
obtained with MAVID; an example is shown in Figure~3. In 
this figure we see that MAVID produced slightly, but systematically better alignments. 
However, because neither algorithm's scoring scheme was optimized, we do not consider 
this figure to indicate which of the two alignment algorithms is better. Rather, it 
represents a proof of principle that our method can be used to identify strengths and 
weakness of different alignment algorithms and evaluate objectively the sequence 
similarity in any given alignment.

\subsection*{Correlations within single translation strings: Shannon informations}

\begin{figure}
\centering
\includegraphics[width=0.68\textwidth]{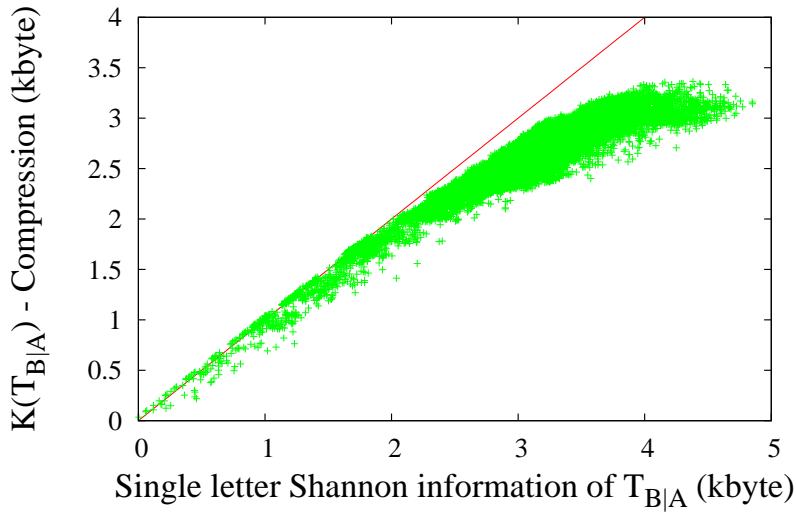}
\caption{ Scatter plot comparing $K(T_{B|A})$ estimated using compression, to the 
   single letter Shannon information $h_1(T_{B|A})$. The diagonal, $X=Y$, is a guide 
   for the eye. Points falling below the diagonal indicate cases where $T_{B|A}$ is 
   not independent and identically distributed, and some letters show strong correlations. The fact that $K(T_{B|A})$
   is slightly larger than $h_1(T_{B|A})$ for low entropy translation strings corresponds
   to the initiation cost for lpaq1 compression, which is $\approx 30$ byte independently
   of the sequence. \blue{The plot shows $\approx 30,000$ pairs taken from all over the animal 
   kingdom.}
\label{fig:shannon-1}}
\end{figure}

In Figure~4 we show compression based conditional complexity 
estimates for animal mtDNA translation strings plotted against the corresponding 
single letter Shannon entropies $h_1$. In the latter, we have not eliminated 
indels, i.e. they are based on the nine letter alphabet $\{0,1,2,3,A,C,G,T,-\}$.
Thus the difference between $K(T_{B|A})$ and $h_1(T_{B|A})$ is entirely based on
correlations, detected by the compression algorithm (in this case lpaq1). 

As $K(T_{B|A})$ goes to zero, the two estimates agree, up to a small initialization cost 
for lpaq1 of $\approx 30$ bytes. The estimates agree because the translation string is 
mostly composed of zeros, with the few substitutions being far apart and weakly correlated. 
For increasing $K(T_{B|A})$, however, the compression algorithm often gives significantly 
lower estimates, thus indicating strong correlations within the translation string. 
More detailed analysis of pairwise correlations (unpublished) suggests that these 
are mostly correlations between letters $A,C,G,T,-$ (\blue{i.e.,} inserts and gaps 
\blue{rather than substitutions}). The fact that indels occur strongly correlated is 
well known \cite{aluru2006}, and is also assumed in most alignment scoring schemes. 

Therefore, if the information encoded in gaps is to be taken into account, it is 
necessary to go beyond the single letter approximation when estimating realistic and
absolute sequence similarities.  Furthermore, taking into account only pairwise letter
correlations would not be sufficient either. This, of course, is not completely new, 
and the most common way to deal with this problem is to simply ignore indels 
\cite{Nei-Kumar2000}. Indeed there seems to exist a wide spread opinion that indels
are not very informative and useful. Whether this is true or whether it just reflects
an inability to deal with this information efficiently is an open question. In any 
case, the most \blue{straightforward} way to deal with it would be based on algorithms 
using data compression. 

\subsection*{Comparison with p-distances: The effect of indels}

\begin{figure}[t]
\centering
\includegraphics[width=0.45\textwidth,angle=270]{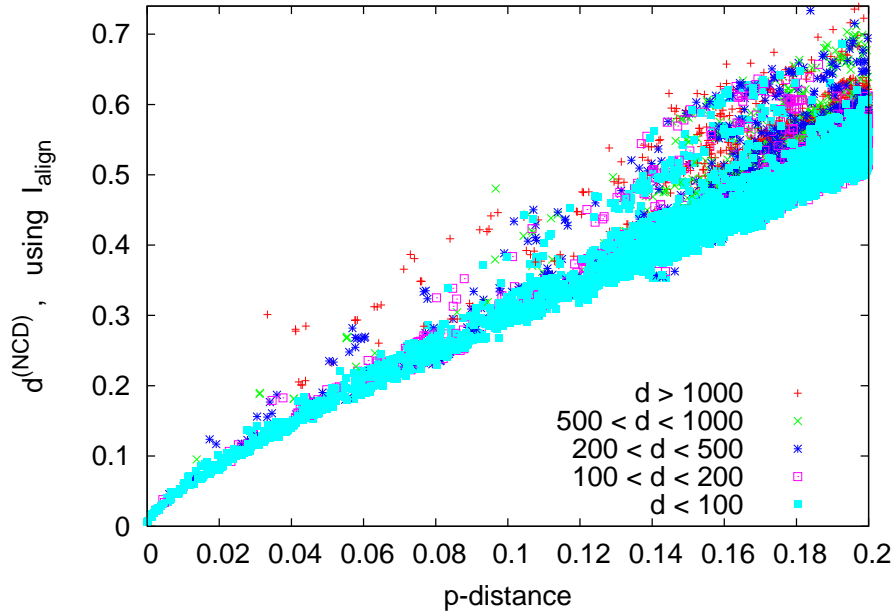}
\vglue 8.mm
\caption{ Scatter plot comparing p-distances $d^{\rm (p)}_{AB}$ to \red{normalized} compression
   distances $d^{\rm(\red{NCD})}_{AB}$ obtained from $I_{\rm align}$. The figure 
   is based on $\approx 10^5$ mtDNA pairs, selected according to criteria discussed in 
   the main text. Different symbols correspond to different length differences 
   $d = |N_A-N_B|$, where $N_A$ and $N_B$ are the original (non-aligned) sequence
   lengths.}
\label{fig:p-distrib}
\end{figure}

A very simple but popular distance measure between sequences (both DNA and amino acids)
is p-distance. It is defined by first removing all indel positions and then counting 
the number of positions where the two sequences disagree \cite{Nei-Kumar2000},
\begin{equation}
     d^{\rm (p)} = {m_{\rm substit}\over M},
\end{equation}
where $m_{\rm substit}$ is the number of observed substitutions and $M$ is the total
number of (non-indel) sites. Since this \blue{quantity} saturates with increasing evolutionary 
distance, a slightly more sophisticated version is the Poisson corrected (PC) p-distance 
\cite{Nei-Kumar2000}, $d^{\rm (p,PC)} = -\log(1-d^{\rm (p)})$. We note that neither $d^{\rm 
(p)}$ nor $d^{\rm (p,PC)}$ take into account the type of substitutions, any information 
contained in indels, or any information contained in internal correlations within the 
translation strings.

Our main interest here is to see which of these three neglected aspects (type of 
substitution, indels, correlations) has the biggest effect. In Figure~5
we show a scatter plot of the \red{normalized} compression distance $d^{\rm(\red{NCD})}_{AB}$, 
estimated via $I_{\rm align}$, against $d^{\rm (p)}_{AB}$, for $\approx 10^5$ pairs 
$(A,B)$ taken from all over the animal kingdom. In order to avoid meaningless alignments, 
we took in each pair only members in the same (sub-)phylum (hexapoda, mollusca, crustacea,
chelicerata, cnidaria, porifera, platyhelminthes, echinodermata) or in the same
(super-)class (mammals, sauropsida, amphibia, actinopterygii). We also eliminated
pairs with $m_{\rm substit}+m_{\rm conserv} <0.9 \min\{N_A,N_B\}$, as we would have otherwise
too many biologically meaningless alignments. Here, $N_A,\; N_B$ are the sequence lengths; this
criterion guarantees that there are not too many insertions into the longer sequence, 
and not too many deletions from the shorter. We found that there is a roughly monotonic 
relationship between $d^{\rm(\red{NCD})}$ and $d^{\rm (p)}$, with occasional, strong, 
deviations. By far the strongest factor leading to these deviations is the difference in 
length of the paired raw (i.e. unaligned) sequences. Nearly all gross outliers in 
Figure~5 correspond to pairs in which one member has a very long 
mitochondrial genome, leading to a large number of indels. 

As we had pointed out in the previous subsection, it is widely believed that indels
are not very informative. We plan to check this \blue{more carefully} in a future publication, 
using a methodology based on a large number \blue{of} quartets \blue{for} sequences similar to the one 
described in the next section.

\subsection*{Comparison with log-det distances: The effect of substitution types}

\begin{figure}
\centering
\includegraphics[width=0.45\textwidth,angle=270]{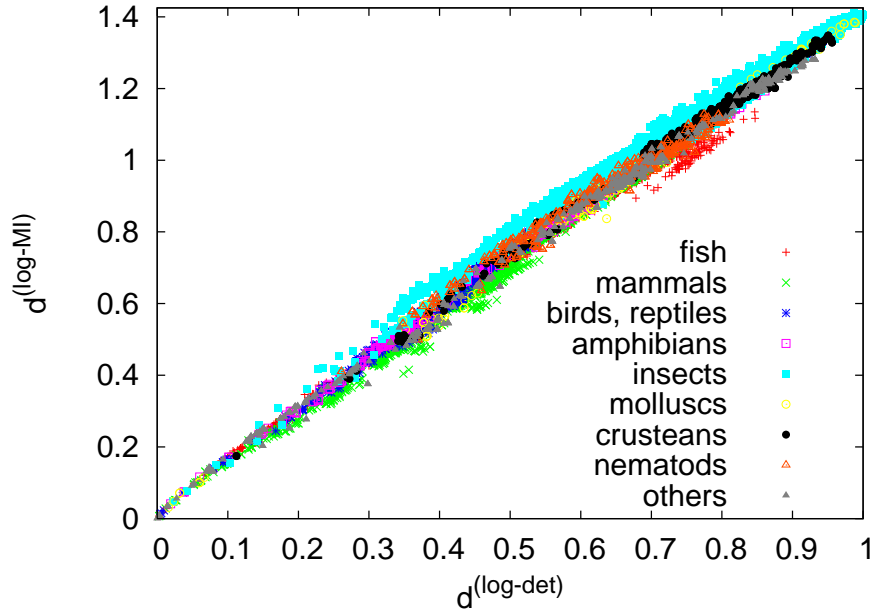}
\vglue 8.mm
\caption{ Scatter plot comparing $d^{\rm (log-MI)}_{AB}$ \blue{(based on single letter Shannon
   entropies)} to $d^{\rm(log-det)}_{AB}$, for $5\times 10^4$ randomly chosen pairs of species.}
\label{fig:logdet}
\end{figure}

Finally, we want to compare our distance metrics $d^{\rm(\red{NSD})}$ and $d^{\rm(log-MI)}$ to 
the log-det distance \blue{$d^{\rm(log-det)}$} given in Eq.~(\ref{logdet}). In order to simplify the 
discussion and to use exactly the same input for all three metrics, we use the same alignment 
algorithm (MAVID) for each pair and delete all indels.  As mentioned above, $d^{\rm(log-MI)}$ 
does not, in general, satisfy the triangle inequality. But this does not preclude it from being 
satisfied in all "typical cases".  To test this we first check whether the triangle inequality 
is actually violated or not in $10^5$ randomly chosen triplets, drawn from the entire animal 
kingdom, with the same selection criteria as in the previous subsection.  Note that due to the 
omission of indels none of these "distances" actually have to satisfy the triangle inequality. 
Indeed, we found 11 violations for the log-det distance, and none for either of the MI-based 
distances.

Next we tried to check whether $d^{\rm(log-MI)}$ is at least approximately additive.
Since we do not have the true evolutionary distances, we take $d^{\rm(log-det)}$
as a proxy. In Figure~6 we plot $d^{\rm(log-MI)}_{AB}$ against
$d^{\rm(log-det)}_{AB}$ for $50,000$ random pairs. We see that:
\begin{itemize}
\item Roughly, the dependence is linear. Thus, to the extent that $d^{\rm(log-det)}$
is linear, $d^{\rm(log-MI)}$ is too. Thus it should not be affected by long branch
attraction. This is in contrast to $d^{\rm(\red{NSD})}$ which -- when plotted against 
$d^{\rm(log-det)}$ -- is strongly non-linear (data not shown).
\item On a finer scale, one sees several deviations. The most conspicuous, perhaps, 
is that insects (hexapods) are systematically above the main curve. This is due to 
the strong compositional bias in most insects, where C/G is underrepresented 
compared to A/T. This reduces the entropies of individual sequences. At the same
time, however, substitution rates involving C and G are not as suppressed.
As a consequence, the ratio $I/K$ is enhanced compared to other phyla, and 
$d^{\rm(log-MI)}$ is increased. This is a desirable effect. It is well known 
\cite{Steel1993} that similar compositional bias can make two sequences look 
more closely related, even if they are not closely related evolutionarily. While 
this applies fully to $d^{\rm(log-det)}$, the effect is at least smaller for 
$d^{\rm(log-MI)}$.
\item For intermediate distances ($0.2 < d^{\rm(log-det)} < 0.5$), many mammals 
are below the main line. In particular, consider the two pairs well below it at 
$d^{\rm(log-det)} \approx 0.35$. Both involve the spectacled bear (Tremarctos 
ornatus) and another Ursinae species. For whatever reasons, these two translation
strings contain an unusually large ratio of transitions to transversions that would otherwise
only \blue{be} typical for much more closely related species. This reduces the information content when compared
to unbiased substitutions with the same total frequencies. At the same time, the 
individual sequences are not very strongly biased. Thus $d^{\rm(log-MI)}$ is reduced, 
but $d^{\rm(log-det)}$ \blue{is not} -- since it is only weakly dependent on the detailed 
substitution rates. Again we claim that this favors  $d^{\rm(log-MI)}$ 
over $d^{\rm(log-det)}$.
\end{itemize}

A clear decision whether this is indeed true can only be made by detailed 
comparison of phylogenies predicted on the basis of these metrics with the 
true phylogenies. Since the latter are of course unknown, we take inferences 
made in the literature as proxies. Our detailed strategy is the following:
\begin{enumerate}
\item
We first choose $10^7$ random quadruples from all over the animal kingdom. 
We use the same taxonometric restrictions, to avoid too many pairs which cannot 
be meaningfully aligned. Thus each quadruple (or ``quartet") contains only 
species from the same (sub-)phylum or the same (super-)class, respectively. 
We also used the same cut on the number of indels, in order to eliminate false 
alignments.
\item
For each quartet, we find the topologies suggested by each of the three metrics, and 
count the number of cases where two metrics disagree. This gave 185543 quartets
(1.9\%) where $d^{\rm(log-MI)}$ and $d^{\rm(\red{NSD})}$ disagree, 429386 quartets
(4.3\%) where $d^{\rm(\red{NSD})}$ and $d^{\rm(log-det)}$ disagree, and 380487 quartets
(3.8\%) where $d^{\rm(log-MI)}$ and $d^{\rm(log-det)}$ disagree.
\item
For each quartet we compute a significance $S$ with which the suggested topology 
is actually preferred. This significance is explained in detail in the supplementary
information. It involves both the amount by which the four-point condition is violated, 
and the relative length of the central edge, if the data are approximated by an 
additive tree. For each pair of metrics we then pick the quartets for which the 
metrics disagree most significantly (as measured by the sum of the two significances). 
Actually, we do not strictly choose the worst disagreements, as they would cluster 
within a few taxa and we want our results to represent as much of the entire animal 
kingdom as possible. As such, we take relatively more quartets in taxa which are 
underrepresented in \red{GenBank}, and we reject quartets (not entirely systematically) 
if three of the four species had already appeared in many selected quartets. 
\item In this way we selected 129 ``worst" disagreements between $d^{\rm(log-MI)}$
and $d^{\rm(log-det)}$, and 129 ``worst" disagreements between $d^{\rm(\red{NSD})}$ 
and $d^{\rm(log-MI)}$. For reasons that will become \blue{clear later}, we did not select 
worst disagreements between $d^{\rm(log-det)}$ and $d^{\rm(\red{NSD})}$, except for a few 
cases. For each of these worst cases we searched the literature and established the ``correct" 
topology. Details are again given in the supplementary material, Tables S1 to S9.
\end{enumerate}

\begin{table*}
\begin{center}
\begin{tabular}{|r|r|r|} \hline
         &    $d^{\rm(log-MI)}$ versus $d^{\rm(log-det)}$ & $d^{\rm(log-MI)}$ versus $d^{\rm(\red{NSD})}$  \\ \hline
  first agrees  &        106             &          57    \\ 
  second agrees &         17             &          42    \\
  neither       &          3             &          24    \\
  undecided     &          3             &           6    \\   \hline
\end{tabular}
\caption{Number of quartets for which each pairwise metric produces a topology that agrees 
   better with that found in the literature.  The quartets  examined are among those for 
   which the disagreement between the two metrics is quantitatively the worst. We note that, 
   when compared to $d^{\rm(log-det)}$, $d^{\rm(log-MI)}$ produces a topology that agrees 
   with literature much more often. ``Neither" indicates the case where neither metric 
   produces a topology that agrees with the current literature. ``Undecided" indicates that 
   it is not possible to establish a 'correct' topology on the basis of current literature.}
   \label{table_1}
\end{center}
\end{table*}

The final results of this are summarized in Table~1. They clearly indicate that the log-MI 
metric is vastly superior the log-det distance, in spite of the latter's superior theoretical 
foundations. This is at odds with the fact that the log-MI metric is not \blue{a} proper 
distance, and does not, in any reasonable model, satisfy the four-point condition 
(Eq.~(\ref{4-pt})). The reason obviously is that $d^{\rm(log-MI)}$ takes into account, in 
an optimal model-independent manner, compositional details that $d^{\rm(log-det)}$ does not.
The comparison between $d^{\rm(log-MI)}$ and $d^{\rm(\red{NSD})}$ is much less clear. One might have 
expected that the strong non-additivity of $d^{\rm(\red{NSD})}$ makes it unsuitable for this sort 
of phylogenetic application. But this is not so clear; $d^{\rm(log-MI)}$ is only marginally 
better. \blue{ This seems surprising, but a possible reason for it will be given in the discussion.} 

Before moving on \red{we highlight} a few notable observations about our quartet 
analysis. Previously, we pointed out that the spectacled bear (T. ornatus) is anomalous either in 
$d^{\rm(log-det)}$ or in $d^{\rm(log-MI)}$.  Indeed, it appears twice in Table S1, and both times 
$d^{\rm(log-MI)}$ gives the correct grouping. A similar anomaly is seen in Figure~6 
for fish (actinopterygii) at $d^{\rm(log-det)} \approx 0.77$. Most of these correspond to Albula 
glossodonta (GenBank AP002973) paired with other fish. Table S2 shows that for most of these pairs 
the log-MI distance gives a better estimate.

We find that discrepancies between $d^{\rm(log-det)}$ and $d^{\rm(log-MI)}$ are very unevenly 
distributed over the taxa.  While we found no disagreements in the chaetognatha, there are a 
large number in the nematods, most favoring $d^{\rm(log-MI)}$. Indeed, it seems that the \blue{nematod}
phylogenetic tree constructed using $d^{\rm(log-det)}$ would be {\it systematically} different 
\blue{from} the tree constructed using $d^{\rm(log-MI)}$ and other analyses.

It is well known \cite{Lake1994,Lockhart1994} that the log-det distance is additive only when 
\blue{the} evolutionary rate is constant over all sites.  One can argue that an analysis that 
does not distinguish sites with different evolutionary speeds is not fair to $d^{\rm(log-det)}$.  
In response we put forth the following three points: (i) The main problem with 
$d^{\rm(log-det)}$ does not seem to be a lack of additivity, but rather insufficient attention 
to the specific types of substitution; (ii) Inhomogeneities in the evolutionary speed should 
affect not only the log-det distance, but most other distance measures as well.  Specifically we 
cannot see why it should not negatively affect $d^{\rm(log-MI)}$ too; (iii) \blue{Similarly}, 
analyzing sites with different speeds separately should improve the results for any distance 
measure -- as long as it can be done unambiguously, without too much effort, and without reducing 
the amount of usable data excessively. In view of the last three caveats we believe that ``naive" 
analyses, such as the one presented above, have and will continue to have their merits.

\subsection*{\blue{The full picture: Comparison of several distance metrics}}

\blue{ So far we have only compared in detail quartet classifications based on log-det distances and 
on single letter Shannon MI. We have used Shannon MI because its estimation is less ambiguous
than compression based MI estimation, and because it uses {\it exactly} the same input ---
the base substitution frequency matrix after removing indels --- as the log-det distance. 
But our tenet is, of course, that compression based estimates should be superior as long as they
use the information about indels efficiently. In addition to the log-det distance, there are several 
measures that are often used. In this subsection we make several pairwise comparisons similar to 
the one made in the previous subsection. But we restrict ourselves to mammals, as these have the 
best understood phylogeny, we expect the least numbers of errors in the literature classification.}

\blue{In this subsection we compare MI based distances will be with the log-det and with both
versions of the Kimura distance (Eqs.~(S8,S9)) discussed in the supplementary 
material. We do not present all possible combinations, as this would produce a huge matrix. Instead, 
we focus on a subset of the distance measure pairs, but we claim that this subset 
is large enough to present a clear overall picture.} 

\blue{Results are shown in Table 2. As mentioned above, we analyzed only mammals for this, but we looked 
at {\it all} possible quartets. Our criteria for identifying the ``worst" disagreements is 
the same as in the previous subsection. Each comparison is based on at least 60 disagreeing quartets. 
In this table, ``Kimura$_1$" and ``Kimura$_2$" refer to Eqs.~(S8) and (S9) in the supplementary material, 
respectively; ``Shannon nolog-MI" stands for $d^{(NSD)}$ 
(Eq.~(\ref{NSD})), ``Shannon log-MI" stands for the logarithmic version of the Shannon distance 
(Eq.~(\ref{logdist}), right hand side), ``transl. string, nolog" stands for $d^{(NCD)}$ (Eq.(\ref{NCD})
with the MI estimated via alignment), ``transl. string, log" stands for its logarithmic version 
(Eq.(\ref{logdist}), left hand side), ``XM compression, nolog" stands for $d^{(NCD)}$ 
with  the MI estimated via concatenation and compression with XM, and ``XM compression, log"
stands for its logarithmic version.} 

\begin{table}[h]
\begin{center}
\begin{tabular}{|l@{ : }l|r@{ : }r@{ : }r@{ : }r|} \hline
   type$_1$              & type$_2$               & first & second  & neither & undecided  \\ \hline

   Shannon log-MI        &  log-det               &  64   &  2      &   0     &     4    \\
   Shannon log-MI        &  Kimura$_1$            &  33   & 24      &   4     &     1    \\
   Shannon log-MI        &  Kimura$_2$            &  51   &  6      &   4     &     0    \\
   Kimura$_1$            &  Kimura$_2$            &  48   &  7      &   3     &     3    \\
   transl. string, nolog &  Shannon log-MI        &  40   & 14      &   6     &     1    \\
   XM compression, nolog &  Shannon log-MI        &  40   & 13      &   0     &     2    \\
   XM compression, nolog &  transl. string, nolog &  52   & 12      &   0     &     1    \\
   Shannon nolog-MI      &  Shannon log-MI        &  53   & 29      &  11     &     7    \\
   transl. string, nolog &  transl. string, log   &  23   & 35      &   2     &     3    \\
   XM compression, nolog &  XM compression, log   &  54   &  3      &   2     &     1    \\ \hline \hline

\end{tabular}
\end{center}
\caption{\blue{Pairwise comparisons between different distance measures for complete mammalian mtDNA. 
   Compared are the abilities of $d^{(\rm type)}$ to correctly classify a large number of quartets. 
   First, the topologies of the quartet trees obtained with two distances $d^{({\rm type}_1)}$ 
   and $d^{({\rm type}_2)}$ are computed. The quartets with ``worst" disagreements are then looked 
   up in the literature. Based on the literature consensus it is decided which of the two topologies 
   is correct -- unless both are wrong, or no consensus can be arrived at due to non-existent or
   conflicting literature. The four numbers in the columns 3 to 6 are the number of cases in which (1) 
   the distance measure $d^{({\rm type}_1)}$ predicted the correct topology, (2) $d^{({\rm type}_2)}$ 
   predicted the correct topology, (3) none of them did, and (4) no decision is possible. }}
\end{table}

\section*{Discussion}

In the present paper we have not presented any detailed application \blue{to a specific open
phylogenetic problem. We also have not considered larger phylogenetic trees, in view of the 
imperfections of all existing distance based tree reconstruction algorithms. Instead, we have 
concentrated on quartets, since there we can obtain high statistics and the inference of the 
tree from a given distance matrix is trivial. Also, for the most detailed numerical comparison 
we have concentrated on Shannon information based methods, rather than on compression based 
methods for estimating MI.} The reason is simply that we desired a comparison with 
other methods (mainly the log-det distance) which is as straightforward and unambiguous
as possible. Indeed, it is trivial to replace Eq.~(\ref{logdet}) by 
Eqs.~(\ref{NSD}),~(\ref{I-shannon}),~(\ref{h-shannon}). In this way we hope to have the best 
chance to convince even skeptical readers that mutual information based distance measures are 
useful in sequence analysis. 

\blue{ We have also presented similar -- but less complete -- analyses based on large numbers of 
random quartets for (at least partially) compression based algorithms and have demonstrated 
that distances based on data compression give even better phylogenies. Indeed, from Table 2 
we can draw a number of conclusions:}
\begin{itemize}
\item \blue{All versions based on MI are better than any version not based on MI.}
\item \blue{Kimura$_1$ (based directly on the log-likelihood of the data with respect to the Kimura model) 
seems better than the conventional Kimura$_2$, which just estimates the total number of substitutions. 
This supports our suspicion that counting transitions and transversions with the same weight is not
a good strategy.}
\item \blue{Nevertheless, $d^{({\rm Kimura}_1)}$ does worse than $d^{(\rm NSD)}$, as expected: As we 
point out in the supplementary material, the log-likelihood given a model $M$ is essentially a coarse 
grained MI, where different substitutions are lumped together (resp., the probabilities predicted 
by $M$ replace the true observed probabilities). It would be hard to see why this should give 
superior results, given the ease and robustness with which single letter Shannon entropies can be 
estimated.}
\item \blue{Within the class of MI based distances, those which do not neglect indels seem 
systematically better.}
\item \blue{Among the latter, distances based on $I_{\rm compr}$ do better than those based on 
$I_{\rm align}$.  This is surprising, as we saw that $I_{\rm align}$ is for mammals systematically 
larger (and thus supposedly better) than $I_{\rm compr}$.}
\item \blue{Logarithmic transformation of MI based distances seems to give mixed results. It improves the
distances slightly for Shannon MI and for $I_{\rm align}$, but it has very negative effects when used
with $I_{\rm compr}$ based on XM. We conjecture that this reflects two sides of the logarithmic 
transformation for distantly related pairs. On the one hand, it largely eliminates {\it systematic} 
errors due to deviations from metric additivity (the Felsenstein phenomenon). On the other hand it 
amplifies noise. To illustrate this, we discuss in the supplementary material a quartet where 
both the original Shannon MI based distances and their log-transformed versions give wrong results,
but for opposite reasons. We speculate that the detrimental effect dominates for $I_{\rm compr}$,
because MI estimation by compression is more noisy (due to the less systematic way that present 
state-of-the-art compression algorithms work) than $I_{\rm align}$.}

\end{itemize}

\blue{Thus, contrary to wide spread opinion, information about indels can be directly used for 
phylogenetics, even without any detailed model for how they were generated.
A more detailed presentation of these data and their implications will be given elsewhere.}

We believe that so far we have only scratched the true potential of (algorithmic)
information theory for sequence analysis.
Several generalizations and improvements are feasible and are listed below: 

(1) Use more efficient encodings of the translation string. For
instance, we only used the letters $A'_i$ and $T_{B|A,i}$ to reconstruct
$B'_i$, but one could also use in addition $A'_{i-1}, B'_{i-1}$, and/or
$T_{B|A,i-1}$.

(2) Use local alignments instead of global ones. In a local alignment
between sequences $A$ and $B$, large parts of $B$ are not aligned with
$A$ at all and are encoded without reference to $A$. Only the aligned
parts give information from $A$ that can be used to recover $B$.
Before making the jump from global to local alignments, an
intermediate step would be  a ``glocal'' alignment tool such as
shuffle-lagan (``slagan'') of \cite{brudno2003gaf}. 

(3) Construct objective measures based on information theory for the
quality of multiple alignments.  A straightforward measure is the
information about sequence $C$ obtained from aligning it
simultaneously with $A$ and $B$. Assume e.g. that the sequences $A$
and $B$ are much more similar to each other than either $A$ and $C$ or $B$ and $C$ 
(as for human, chimpanzee, and chicken).  In order to measure the MI between
chicken and the primates, one could first align $A$ and $B$ and then align,
in a second step, $C$ to the fixed alignment $(A,B)$.

\section*{Conclusions}

At present, biological sequence analysis is heavily based on the concept 
of alignment. There exist proposals for alignment-free approaches, and it 
has been suggested that they will become more and more important as 
more sequence data become available \cite{vinga2003}. To us it seems 
an open question whether alignment-free algorithms for sequence comparison
will become widely used, whether they will eventually displace 
alignment-based algorithms, or whether both approaches will merge into a unified
approach. We hope that we have shown with the present work that an 
amalgamation of both methods (alignment-based and alignment-free) is possible.
More precisely, by showing that mutual informations between two sequences can 
be easily estimated from global alignments, we have established a direct link
between sequence alignment, \blue{Shannon information theory}, and methods based on 
data compression and Kolmogorov information theory. Technically, we have dealt only with pairwise global 
alignment, but at least the basic concepts should have much wider applicability. 

From another point of view, the present paper deals 
with the basic notion of {\it parsimony}. In bioinformatics (and in phylogeny
in particular) maximal parsimony in dealing with several objects is often taken 
as synonymous to minimal number of changes needed to go from the description of 
one object to the description of another. This is most clearly formulated in the 
so-called ``maximum parsimony method" of distance-free phylogenetic tree 
construction \cite{Nei-Kumar2000}, but it also underlies the concepts of p- and 
log-det distances. However, the invention of the Morse alphabet in the nineteenth century, 
and the theoretical works by Shannon, Kolmogorov, and others in the middle of 
the last century might cast some doubt on it. It is Rissanen's {\it minimum 
description length principle} \cite{Rissanen,Rissanen2}, however, that 
makes this view obsolete today. Instead of paying attention to the {\it 
number} of changes, one should pay attention to the {\it information} needed to 
encode these changes. We call this ``true parsimony". In this sense, the maximum 
parsimony method does not really aim for maximal true parsimony. On the other 
hand, likelihood based and Bayesian methods do aim for true parsimony, but at the 
cost of depending on explicit models. One goal of the present paper is to show 
how true parsimony can be measured in \blue{less model dependent} ways and how maximum true 
parsimony can be achieved to various degrees of approximation. Moreover, even 
the crudest approximation -- based on \blue{MI obtained via} single-letter Shannon 
entropies, with all information about indels discarded -- can lead to important practical 
improvements.


\bibliography{mm1_clean}

\end{document}